\documentclass[12pt]{iopart}

\usepackage[backend=biber,style=numeric-comp,sorting=none,doi=false,isbn=false,url=false, date=year]{biblatex}
\bibliography{biblio}

\usepackage{iopams}

\expandafter\let\csname equation*\endcsname\relax
\expandafter\let\csname endequation*\endcsname\relax
\usepackage{amsmath}
\usepackage{empheq}

\renewcommand{\phi}{\varphi}
\newcommand{\abs}[1]{\left \arrowvert #1 \right \arrowvert}
\AtEveryBibitem{%
  \clearfield{issn} 
  \clearfield{doi} 
  \clearfield{month}
   \clearfield{day}
    \clearfield{issue}
  \ifentrytype{online}{}{
    \clearfield{url}
  }
}

\begin{document}

\title[Study of longitudinal fluctuations of the SK model]{Study of longitudinal fluctuations of the Sherrington-Kirkpatrick model}

\author{
Giorgio Parisi$^1$,
Leopoldo Sarra$^2$\footnote{Present address: Max Planck Institute for the Science of Light, Staudtstrasse 2, 91058 Erlangen, Germany}, 
Lorenzo Talamanca$^2$\footnote{Present address: Ecole Polytechnique F\'ed\'erale de Lausanne, SV IBI UPNAE, CH 1015 Lausanne}
}
\address{$^1$ Dipartimento di Fisica, Sapienza Universit\`a di Roma, INFN – Sezione di Roma 1,
and CNR-NANOTEC UOS Roma,  P.le A. Moro 2, I-00185, Rome, Italy}
\address{$^2$Dipartimento di Fisica, Sapienza Universit\`a di Roma, P.le A. Moro 2, I-00185, Rome, Italy}

\eads{
\mailto{giorgio.parisi@roma1.infn.it},
\mailto{leopoldo.sarra@mpl.mpg.de},
\mailto{lorenzo.talamanca@epfl.ch}
}
\vspace{10pt}
\begin{indented}
\item[] \today
\end{indented}

\begin{abstract}
We study finite-size corrections to the free energy of the Sherrington-Kirkpatrick spin glass in the low-temperature phase where replica symmetry is broken.
We investigate the role of longitudinal fluctuations in these corrections, neglecting the transverse contribution. 
In particular, we are interested in the value of exponent $\alpha$ that controls the finite volume corrections: $\alpha$ is defined by the relation $f-f_\infty\sim N^{-\alpha}$, $N$ being the total number of spins. 
We perform both an analytical and numerical estimate of the analytical result for $\alpha$.
From both the approaches, we get the result: $\alpha\approx0.8$.
\end{abstract}

%
%
%
%
%


\section{Introduction}
The behavior of finite-size corrections to the Sherrington-Kirkpatrick (SK) spin glass in the low-temperature phase \cite{infiniteranged} has been debated for several years. 
For example, we know very precise results can be found in \cite{billoire}: within a numerical approach, he calculated the finite volume corrections at a finite temperature and found $f-f_\infty  \sim N^{-2/3}$. 
The same value of the exponent has been found in \cite{martin1} and \cite{martin2}, studying the $N$ dependence of the ground state.
 The theory of the fluctuations around the mean field solution is quite complex and it is very difficult to perform an estimate of the analytic result, no final value has been found so far. 
Indeed, fluctuations in the replica space give rise to three different kinds of contributions, which are very complicated to handle analytically.

As a starting point to better understand the behavior of fluctuations, we restrict ourselves to consider only a part of them, the so-called longitudinal fluctuations. 
They are defined as those fluctuations that satisfy the structure of the RSB ansatz. In this case, the correlations are functions of only two variables (three variables are needed for the full transverse computation).
Since the longitudinal fluctuations can be calculated directly from the full-RSB framework, it is easier to obtain a prediction for the behavior of their contribution.  
The first attempt in this direction was made in \cite{biscari}.
Studying only longitudinal fluctuations, it was found that $f-f_\infty  \sim N^{-3/4}$ by employing an oversimplified scaling argument.
In this work, following \cite{ferraro}, where a semi-regularized propagator has been attained, we pursue both a fully analytic and a numerical estimate of the analytic result, without using these simplifications.

In Sec. \ref{sec:longitudinal}, we study the structure of longitudinal fluctuations and find the diagrammatic structure of the theory.
Then, in Sec. \ref{sec:scalanas}, we study how the diagrams of our loop expansion behave in a convenient Fourier space at various orders and find our first result.
In Sec. \ref{sec:regolarizza}, we study the effect of a different possible regularization, and we obtain the same result in this new way.
Then, in Sec. \ref{sec:cancellazioni}, we focus on some peculiar relations among the vertices of the theory.
Finally, in Sec. \ref{sec:numeric}, we conduct the numerical estimate of the analytic results.
We choose a convenient observable and implement a Metropolis algorithm.
We analyze our data from two different perspectives:
the first is a more direct one that yields a very similar result to the one of Sec. \ref{sec:scalanas}, the other gives a slightly different prediction in a more tortuous way.

\section{Longitudinal fluctuations}
\label{sec:longitudinal}
We consider the infinite-ranged model proposed by Sherrington and Kirkpatrick \cite{solvablemodel}.
Its solution was found in the so-called full-RSB framework \cite{orderpar} and formalized in a mathematical way in \cite{panchenko}.
In particular, we know the explicit solution only near the critical temperature \cite{de2006random}.
In this region we can write
\begin{equation}
\label{eq:free_energy_discrete}
    f = -\left(\frac{1}{\beta} \log2+ \frac{\beta}{4} \right)  + \frac{1}{N}\log 
    \int d Q_{ab} e^{-N \mathcal{L}_\text{Tr} [Q_{ab}]},
\end{equation}
with 
\begin{equation}
\label{eq:lagrtrunc}
    \mathcal{L}_\text{Tr}=-\left( \frac{\tau}{2}\text{Tr}Q^2 + \frac{1}{6}\text{Tr}Q^3+ \frac{1}{12}\sum \left(Q_{ab}\right)^4 \right)
\end{equation}
and $\tau=(T_C-T)/T$.
It has been found that, for the $N\to \infty$ limit, the quantity in eq.(\ref{eq:free_energy_discrete}) converges to the following limit:
\begin{equation}
\label{eq:f_infty}
    f_\infty=-\left(\frac{1}{\beta} \log2+ \frac{\beta}{4} \right) +\mathcal{L}_\text{Tr} [Q^*_{ab}].
\end{equation}
The  matrix $Q^*_{ab}$ is given by
\begin{equation}
 Q^*_{ab}= \underset{Q_{ab}}{\text{argmax}} 
    \left( e^{-N \mathcal{L}_\text{Tr} [Q_{ab}]} \right).
\end{equation}
This matrix can be parametrized with a function $\bar q(x)\in [0,1]$, with $x\in [0,1]$.
Near the critical point, we have that $\bar q(x)$ is given by: 
\begin{equation}
\label{eq:qbar}
    \begin{cases}
    \Bar{q}(x)=q'(0) x  & x \le x_1=1-\sqrt{1-4\tau} \\
    \Bar{q}(x)=q(x_1)   &  x > x_1.
    \end{cases}
\end{equation}
We can split the original measure $d Q_{ab}$ in  (\ref{eq:free_energy_discrete}) into two orthogonal contributions, longitudinal ones, $Q^L_{ab}$,  which are defined as those that we can write with $q(x)$, and transverse ones, $Q^T_{ab}$.
We have $d Q_{ab} = d Q^L_{ab} d Q^T_{ab}$. 
In what follows, we will only consider fluctuations over $\bar q(x)$, which are by definition the longitudinal fluctuations.
In this case, we  can write the free energy of the system near the critical temperature as \cite{de2006random}
\begin{equation}
\label{eq:Freee}
    f =  -\left(\frac{1}{\beta} \log2+ \frac{\beta}{4} \right) + \frac{1}{N}\log \int \delta q e^{-N\mathcal{L}[q(x)]},
\end{equation}
with the integral restricted to physical $q(x)$, i.e. $\Dot{q} \geq 0$.
If we do not require this condition, the integral will not be defined because the functional is not one-side bounded \cite{Auffinger2015}.

With our assumptions, eq. (\ref{eq:lagrtrunc}) becomes
\begin{equation}
\label{eq:L_q}
    \mathcal{L}[q(x)] =- \int_0^1 \left(\frac{\tau}{2} q(x)^2
        - \frac{1}{6}\left(x q(x)^3
        + 3q(x)^2\int_x^1 q(y) dy\right)
        + \frac{1}{12}q(x)^4 \right)dx.
\end{equation}
We  define  the fluctuation $ \phi(x)=q(x) - \Bar{q}(x)$ and perform a loop expansion around $\Bar{q}(x)$ \cite{mezard1987spin}. 
We notice that the condition $\dot q (x)>0$ does not enter perturbatively in the $\phi(x)$ in the region outside the plateaux; however,  $\Dot{ \Bar{ q}}(x)=0$ implies $\dot \phi (x)=0$ and therefore we take a constant $\phi(x)$ in the plataux.

After some algebra, we obtain
\begin{align}
\label{eq:effective_theory}
    f-f_\infty =\frac{1}{N}\log\left(\int \delta \phi e^{-\frac{1}{2}\mathcal{K}(\phi(x)\phi(y)) - \frac{1}{\sqrt{N}} \mathcal{V}_3(\phi(x)\phi(y)\phi(z)) - \frac{1}{N} \mathcal{V}_4(\phi(x)\phi(y) \phi(z)\phi(w))   }\right),
\end{align}
where the quadratic term is given by
\begin{equation}
  \mathcal{K}(\phi(x),\phi(y))=-\int \left(\bar q(x)\theta(y-x)+\bar q(y)\theta(x-y) \right)\phi(x)\phi(y)dx dy.  
\end{equation}

\noindent Furthermore, this theory consists of a three-legged vertex
\begin{equation}
\label{eq:tk}
    \mathcal{V}_3(x,y,z)  =
           \delta(x-y)\theta(z-x) 
        + \delta(y-z)\theta(x-y)
        + \delta(z-x)\theta(y-z)
\end{equation}
and a four-legged one,
\begin{equation}
\label{eq:pk}
     \mathcal{V}_4(x,y,z,w)=-2\delta(x-y)\delta(y-z)\delta(z-w).
\end{equation}

\noindent The propagator of this theory has already been found in \cite{ferraro}:
\begin{equation}
  G_0(x-y) =-2 \delta'' (x-y),
\end{equation}
with the property that the propagator $G_0$ is the inverse (in the sense of integral operators) of the kernel $\mathcal{K}$.

We can study our problem in the Fourier space of the $x$ variable. We use the following prescriptions:
\begin{align}
    &\widetilde{G}(p) \sim 2 p^2\text{ for the lines},\\
    &\widetilde{V_3} (p_{1},p_{2}, p_{3})\sim 1/p_{1} +1/p_{2}+1/p_{3}  ,\quad \sum p_i=0 \text{ for the three-leg vertex}, \\
    &\widetilde{V_4} (p_{1},p_{2},p_{3},p_{4})\sim 1,\quad \sum p_i=0 \text{ for the four-legged one}.
\end{align}

\section{Scaling argument}
\label{sec:scalanas}
In what follows, we look for the exponent $\alpha$ defined as
\begin{equation}
    f-f_\infty  \sim N^{-\alpha}.
\end{equation}
The naive perturbative theory in $1/N$ produces divergent terms signaling that the coefficient of the $1/N$ is infinite and that the exponent $\alpha$ is less than $1$.

\begin{figure}
\includegraphics[width=0.95\linewidth]{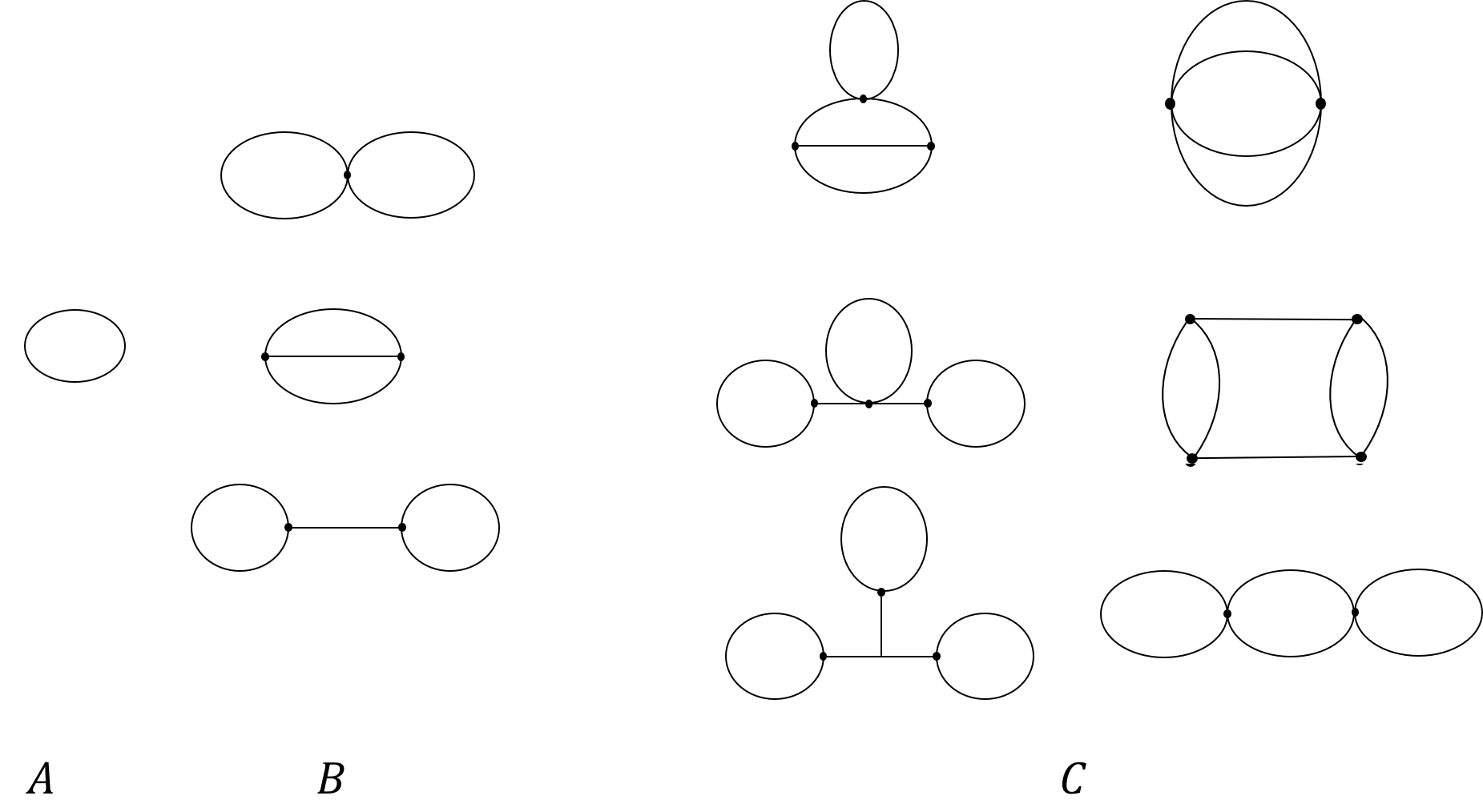}
\caption{Diagrams of the theory of longitudinal fluctuations at one loop ($A$), two loops ($B$) and three loops ($C$).} 
\label{fig:exspectre}
\end{figure}

As usual in field theory, we regularize the theory with a cutoff $\Lambda$ at high impulses and only after having resummed the perturbative expansion we study the limit $\Lambda \to \infty$.
Using these prescriptions, we can easily calculate the divergence of the diagrams in figure \ref{fig:exspectre}.

In fact, we obtain by a simple computation that the contribution to $Nf_\Lambda$ of the one-loop diagrams diverges as $\Lambda$, all two-loop diagrams give $\Lambda^6/N$, and all the three-loop diagrams give $\Lambda^{11}/N^2$.
We find that, as shown in \ref{app:diagrams}, the general behavior is
\begin{equation}
    \frac{\Lambda^{5n+1}}{N^n},
\end{equation}
with $n=L-1$ and $L$ the number of loops of the diagram.
Therefore, the free energy has the form 
\begin{align}
\label{eq:beginF}
        Nf_\Lambda &\sim C_0 \Lambda + C_1 \frac{\Lambda^6}{N}+
        C_2 \frac{\Lambda^{11}}{N^2}+
        C_3 \frac{\Lambda^{16}}{N^3} +
        \ldots 
        \sim \Lambda g\left(\frac{\Lambda^5}{N}\right),
\end{align}
with $g$ an appropriate function.
Let's suppose that $g$ behaves as a power of its argument.
In order for the intensive free energy to be finite in a non-trivial way, in the limit $\Lambda \to \infty$, we find that we necessarily need
\begin{equation}
    \label{eq:freeresult}
    f \sim \frac{\Lambda}{N} \left(\frac{\Lambda^5}{N}  \right)^{-\frac{1}{5}}\Longrightarrow \quad 
    \alpha = \frac{4}{5}.
\end{equation}
This is our prediction for the behavior of the free energy with a finite-size Sherrington-Kirkpatrick model, if only longitudinal fluctuations existed.
We notice that we did not care about the coefficients of the expansion provided that the limit $\Lambda \to \infty$ of the corrections at fixed $N$ exists and it is non zero. If this is true the precise values of the coefficients is irrelevant if we do not aim to compute the prefactor.

If some order of the expansion completely canceled out, the structure of the previous argument would not change. 
This still holds even if the first order is the one that cancels out.
The only concern we should have in assuming that $g(x)$ behaves as a power, is to ensure that an infinite number of diagrams has non zero coefficients. 
This is a reasonable assumption, as if it were false we would have a sum of a finite number of positive powers of $\Lambda$, and the free energy would certainly diverge.

\section{Regularized propagator}
\label{sec:regolarizza}
In \cite{ferraro}, a regularized propagator was presented for this theory. However, it still contained a Dirac delta contribution.
Here, we propose a new regularization to further reduce the divergence of the propagator. 
To do this, we add to the effective action $\mathcal{L}$ in  (\ref{eq:L_q})  the term
\begin{equation}
    \frac{1}{2} R^2\int_0^1 \left(\frac{d\phi(x)}{dx}\right)^2 .
\end{equation}
With similar calculations to those in \cite{ferraro}, we obtain the equation
\begin{equation}
\label{eq:propagator_STP}
    \Bar{q}'(x) G(x-y) + R^2 G''''(x-y) = -\delta''(x-y).
\end{equation}
We can solve this equation in the Fourier space of the $x$ variable, and obtain
\begin{equation}
        \Bar{q}'(x) \tilde{G}(p) + R^2 p^4 \tilde{G}(p) = p^2 
        \quad \Longrightarrow \quad
        \tilde{G}(p) = \frac{p^2}{\frac{1}{2}+R^2 p^4}.
\end{equation}
Finally, we anti-transform it back.
Using well-known properties of the Fourier transform, we find
\begin{equation}
    G(x)= \frac{1}{2^{\frac{1}{4}}R^{3/2}} e^{-\frac{\abs{x}}{2^{3/4} \sqrt{R}}}
        \left(
    \sin{ \frac{\abs{x}}{2^{3/4} \sqrt{R}} } - \cos{\frac{x}{2^{3/4} \sqrt{R}}}
        \right). 
\end{equation}

Now, as we did before, we study the regularized theory in the Fourier space.
We rescale all integrals so that each $p \to  R^{-1/2} p$ and,
following the same steps shown in \ref{app:diagrams}, we obtain that every $n+1$ loop diagram diverges as
\begin{equation}
    \frac{1}{N^n R^\frac{5n+1}{2}} .
\end{equation}
Again, we find $\alpha = 4/5$.
This is a satisfying consistency check.

\section{A relation across the vertices of the theory}
\label{sec:cancellazioni}
In this section, we show a suggestive relation between the vertices of our theory at tree level.
We start by explicitly calculating the four-point amputated function $ \Gamma^{(4)}$.
We have:
\begin{equation}
\label{eq:doublethreeleg}
    \Gamma^{(4)}=3\int_0^1 \mathcal{V}_3(x,y,u) G(u-v) \mathcal{V}_3(z,w,v) dudv-\mathcal{V}_4(x,y,z,w).
\end{equation}
After integrating by parts, we have a total of nine different contributions:
\begin{itemize}
    \item one term equal to $V_4(x,y,z,w)$,
    \item one term equal to $-\theta (x-y) \delta (z-w) \delta'(z-y)$ and the other three permutations,
    \item one term equal to $\theta(x-y) \theta (z-w) \delta''(y-w)$ and the other three permutations.
\end{itemize}
By pairing a term from the second category and one from the third one, we can  regroup those contributions. 
We apply a  $\phi(x)$ to each leg and integrate, the integration variable being  $x$. 
Using the fact that at tree level
\begin{equation}
    \phi(x) = \int G(x-y)\epsilon(y) dy=-2\int\delta''(x-y)\epsilon(y)dy = -2\epsilon''(x),
\end{equation}
we can solve three of the four integrals we started with, and sum all the four pairs.
So, we find that the term of $\Gamma^{(4)}$ not proportional to $\mathcal{V}_4$ is
\begin{equation}   
-4\int_0^1 \epsilon'(x)\epsilon'''(x)\frac{d}{dx}\left(  \epsilon'(x)\epsilon''(x)  \right)dx.
\end{equation}
We notice that, if the external field had a zero third derivative, the four point connected correlation function would trivially be proportional to the four-legged vertex.
This result is similar to what was realized in \cite{Temesvari2000} and \cite{de1998ward}.

\section{Numerical evaluation}
\label{sec:numeric}
We verify the previous argument through a numerical calculation of the functional integral in eq. (\ref{eq:Freee}).
We notice that:
\begin{equation}
\label{eq:startnumerical}
    \langle \mathcal{L} \rangle_N =
        \frac{1}{Z}\int \delta q(x) e^{-N  \mathcal{L}[q(x)]}\mathcal{L}[q(x)]
    = - \frac{\partial }{\partial N}
         \log \int\delta q(x) e^{-N \mathcal{L}[q(x)]}.
\end{equation}
From this equation we get, as shown in \ref{app:lagrangian},
\begin{equation}
\label{eq:free_energy_double_expert}
f - f_{\infty} \sim \langle \Delta \mathcal{L} \rangle_N ,
\end{equation}
with $\Delta \mathcal{L}\left[q(x)\right] = \mathcal{L}\left[q(x)\right] - \mathcal{L}\left[\overline{q}(x)\right]$.
Therefore, we can evaluate $\langle \Delta \mathcal{L} \rangle_N$ employing a convenient Monte Carlo technique.
Regarding the functional integral on the right hand side of  (\ref{eq:free_energy_double_expert}), we need only consider $q(x)$ functions that are monotonically increasing and constant after a fixed value $x_1$, defined in (\ref{eq:qbar}).
Thus, we divide $q(x)$ in $K$ bins $q_k$ for $x\leq x_1$ and consider a single bin $q_K$ for $x>x_1$.
We sample the phase space with a Metropolis algorithm according to which we vary the function $q(x)$ by changing the value of a single $q_k$, with $q_0$ fixed in zero.
The new value is randomly extracted as
\begin{gather}
q_k^{\text{new}} = \begin{cases}
    0 & k = 0, \\ \nonumber
    \text{uniform in }  [q_{k}-\rho_1(q_{k}-q_{k-1}),q_{k}+\rho_1(q_{k+1}-q_{k})] & 0\neq k \neq K,  \\ \nonumber
    \text{Gaussian with mean $q_K$ and variance $\sigma=(\rho_2 x_1/K)^2$} &k=K \\\nonumber
    \text{ and constrained in $[q_{K-1},1]$} &
    \end{cases}
\end{gather} 
where $\rho_1$ and $\rho_2$ are two appropriate parameters we choose in order to keep the acceptance ratio of the changes we propose above $15\%$.

\begin{figure} 
\centering
\includegraphics[width=0.8 \linewidth]{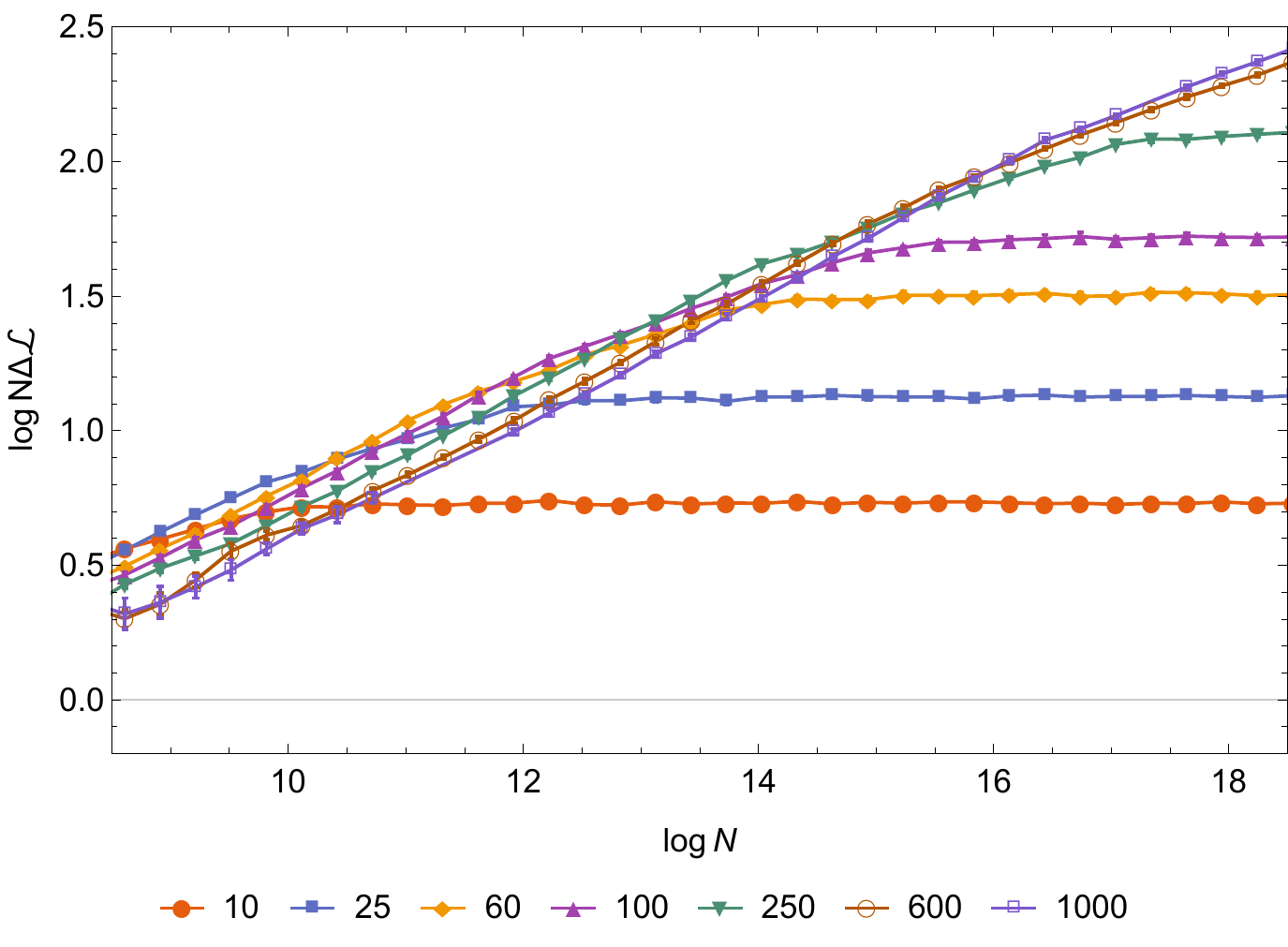}
\caption{Set of curves $N\Delta \mathcal{L}(N)$ at various $K$, reported on log-log plot.}
\label{fig:disegna}
\end{figure}

At fixed temperature $T=0.8$, we run the program for some different values of the discretization $K$ between $10$ and $1000$, namely $\{10$, $ 15$, $ 25$, $ 40$, $ 50$, $ 60$, $ 75$, $ 85$, $ 100$, $ 150$, $ 200$, $ 250$, $ 300$, $ 400$, $ 500$, $ 600$, $ 750$, $ 850$, $ 1000   \}$.
We study the range $N\in [10^5,10^{18}]$.
Each $N_i$ is chosen by doubling the previous one. 
Every time, we ensure to start the measurement after the system reached equilibrium.
We employ $M = 10^5$ Monte Carlo sweeps and average on $S=10$ repetitions. 
Simulations were performed with fixed $K$ and increasing $N$ so that for each repetition we could use a simulated annealing procedure in $N$.
A significant subset of the data is reported in figure \ref{fig:disegna}.  

To extrapolate the behavior of the free energy we need to take two limits,
\begin{equation}
    \lim_{N \to \infty} \lim_{K \to \infty} \langle \Delta \mathcal{L}(K) \rangle_N,
\end{equation}
making sure to take them in the correct order.
We consider only data with $N> 10^9$ as we are interested in the $N \to \infty$ limit. 
For each $K$ we have two different effects:
discretization ones occur for $N>\Bar{N}_K$, for some $\Bar{N}_K$, and they are manifest through the plateau to the right;
further corrections occur for $N<\Hat{N}_K $, for some $\Hat{N}_K \leq \Bar{N}_K$, and cause the mixing of the curves for different values of $K$.
Both effects produce a decrease in $N\Delta\mathcal{L}$. 
We can remove the dependence on $K$ in two ways.

A first approach is to take, for each $N$, the maximum value of $N \Delta \mathcal{L}$.
In this way, it is like we are neglecting the smaller values of $N \Delta \mathcal{L}$ we get due to the effects we described before.
\begin{figure} [ht]
\centering
\includegraphics[height=5.0cm]{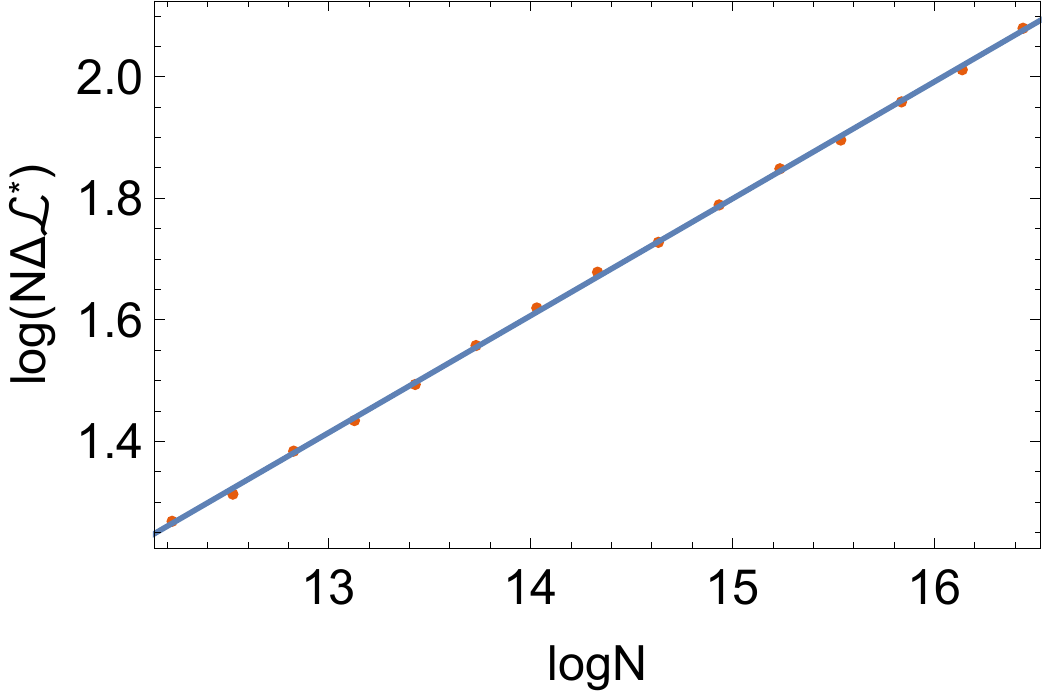} 
\caption{Log-Log plot of the maximum of $N\Delta \mathcal{L}$ as a function of $N$. We find the slope of the line $a=0.192$}
\label{fig:disegnasantanas}
\end{figure}
We plot the data in a log-log scale and find the slope $a$ of the curve, as shown in figure \ref{fig:disegnasantanas}.
We finally find 
\begin{equation}
\label{eq:first}
    \alpha = 1-a = 0.808 \pm 0.004.
\end{equation}
To estimate the errors in our data, we employ the jack-knife method.
We consider the value obtained by each of the $S$ repetitions and we average among sets of five.

\begin{figure}
\centering
\includegraphics[width=0.6 \linewidth]{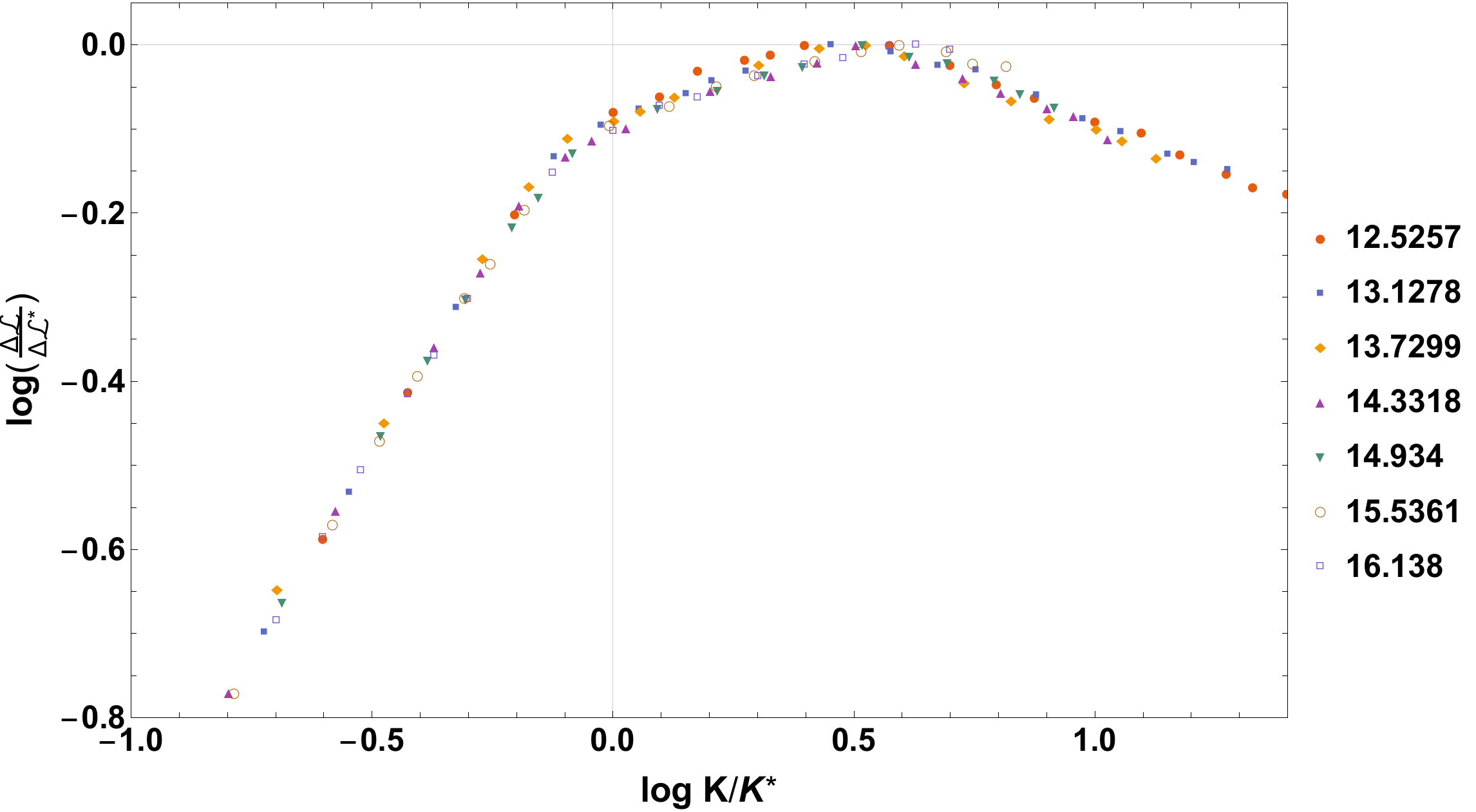}
\caption{Scatter plot that shows all our data correctly scaled. We show various curves $\Delta\mathcal{L}(K)$ at fixed $N$. We report in the legend the $\log_{10}(N)$ of each series.}
\label{fig:scatterplot}
\end{figure}

We can also study our data from another perspective. 
As we said, at a fixed $N$,   $N\Delta \mathcal{L}$ becomes almost constant after some $K=K^*$.
We define the exponents $\beta$ and $\nu$ in the following way:
\begin{align}
\label{eq:lotto}
   & N \langle \Delta \mathcal{L}(K) \rangle_N \sim K^\beta 
    \quad \text{if} \quad \quad K<K^*\\ \nonumber
     &   N \langle \Delta \mathcal{L}(K) \rangle_N \sim N \Delta \mathcal{L}(K^*) 
    \quad \text{if} \quad \quad K>K^*\\ \nonumber
    &   K^*\sim N^\nu.
\end{align}
We are interested in $\Delta \mathcal{L}(K^*)$, i.e. the point where $\Delta \mathcal{L}(K)$ becomes independent of K, so that the order of the limits is respected.
Since
\begin{equation}
      \langle \Delta \mathcal{L}(K^*) \rangle_N \sim   (f-f_\infty)\sim N^{-\alpha},
\end{equation}
we have
\begin{equation}
  \alpha = 1 -\beta \nu.
\end{equation}
For all considered $N$ values we find almost the same $\beta$: the average is
\begin{equation}
    \beta = 0.97 \pm 0.02.
\end{equation}
Fitting the data, we obtained $\nu = 0.189 \pm 0.003$. 
Our result is 
\begin{equation}
\label{eq:second}
    \alpha = 1- \beta \nu = 0.817 \pm 0.006.
\end{equation}
Also in this case to estimate the error for $ \alpha $, we employ the jack-knife method as described before.
This is in agreement with what we found in the previous section.
To see the consistency of the data and the effectiveness of the exponents we defined,  in figure \ref{fig:scatterplot} we show a scatter plot, in which we plot $\Delta \mathcal{L}/\Delta \mathcal{L}^*$ as a function of $K/K^*$. 
This plot clearly shows that the curves $\Delta \mathcal{L}(K)$ overlap almost perfectly. 

The most accurate value from the numerical analysis gives a value of $\alpha = 0.808 \pm 0.004$. However, the error is only statistical: we do not consider systematic errors induced by correction to scaling. The numerical results are thus perfectly compatible with the analytic treatment.

\section{Conclusions}

To sum up, we have taken a step forward towards the understanding of finite-size corrections to the SK model. 
In particular, we obtained a reliable result for the contribution of purely longitudinal fluctuations. 

Firstly, we must confront ourselves with \cite{biscari}.
Our result is not far from theirs.
However, their result is different because they assumed that further order matrices in the expansion are diagonal in the base where the Gaussian matrix is, without having any idea of the repercussions; on the other hand, we tried and managed to keep under control all the approximations we deemed appropriate.
In the numerical analysis in \cite{billoire}  all kinds of fluctuations are relevant.
If longitudinal fluctuations are not the dominant kind, it is reasonable that we found a slower divergence and a higher value of $\alpha$.

In addition, we found the same analytic result $\alpha = 0.8$ with two different regularizations. This is also well consistent with our numerical approaches that yields $\alpha = 0.808 \pm 0.004$ and $\alpha = 0.817 \pm 0.006$.
We also provided a new regularized propagator and a suggestive identity, which can be used in the future to further study the longitudinal fluctuations theory. We hope that a similar computation can be performed for the full theory along the lines of Sec. \ref{sec:scalanas}.

We would like to thank Federico Ricci-Tersenghi and Andrea Crisanti for helpful discussions.
This project has received funding from the European Research Council (ERC) under the European Union’s Horizon 2020 research and innovation program (Grant No. 694925).

\appendix
\section*{Appendix}

\section{Derivation of the analytical result}
\label{app:diagrams}
In general, each order of the loop expansion in  (\ref{eq:effective_theory}) has the same degree of divergence. 
We find that, using the prescriptions given in Sec. \ref{sec:longitudinal}, we can assign to each part of the integral a dimension:
\begin{align}
    [\widetilde{G}(p)]&=2, \\
    \left[\int dp\right]&=1, \\
    [\widetilde{V}_3]&=-1, \\
    [\widetilde{V}_4]&=0.
\end{align} 
If the sum of the dimensions of all the components of an integral is $D$, the diagram diverges as $\Lambda^D$.
Therefore, any diagram with $L$ loops diverges as
\begin{equation}
    D=L+2I-V_3,
\end{equation}
where $I$ is the number of internal lines and $V_3$ is the number of three-legged vertices. 
This result can be verified by counting the powers of the impulses in each integral. 

As we stated previously, each integral carries dimension $1$, and we have an integral for each loop in the diagram. 
Furthermore, each propagator, associated with a line in the diagram, has dimension $2$, each vertex $V_3$ has dimension $-1$ and $V_4$ vertices carry no dimension.
In addition, the following relations hold:
\begin{align}
    4V_4 + 3V_3 = 2I \label{eq:linee}, \\
    I-(V_3+V_4-1)=L.\label{eq:loopz}
\end{align} 
In fact, a closed diagram has $4V_4 + 3V_3 $ legs to connect with lines; since  each line connects two and only two legs, we have (\ref{eq:linee}).
Also, if the diagram has $V=V_3+V_4$ vertices, we necessarily need $V-1$ lines to connect all of them, and any additional line creates a loop, thus (\ref{eq:loopz}).
Each diagram has a weight of $1/N^n$.
We easily find, from (\ref{eq:effective_theory}), that $n=2V_3 + V_4$.
As a consequence, the divergence of the $n+1$ loop is
\begin{equation}
    \frac{\Lambda^{5n+1}}{N^n}. 
\end{equation}

\section{Derivation of the observable}
\label{app:lagrangian}
We consider the average value of $\mathcal{L}$:
\begin{equation}
    \langle \mathcal{L} \rangle_N =\frac{1}{Z}\int \delta q(x) e^{-N  \mathcal{L}[q(x)]}\mathcal{L}[q(x)]=  \frac{\partial }{\partial N}
    \log \int\delta q(x) e^{-N \mathcal{L}[q(x)]}.
\end{equation}
Substituting into the right hand side of the free energy equation (\ref{eq:Freee}), we find
\begin{equation}
  \beta    \frac{\partial }{\partial N} \left(F + \frac{N}{\beta} \log2+ \frac{\beta N}{4} \right) =\langle \mathcal{L} \rangle_N.
\end{equation}
Then, we integrate this equation between $0$ and $N$:
\begin{equation}
    \beta  \left(F + \frac{N'}{\beta} \log2+ \frac{\beta N'}{4} \right)\Bigg\arrowvert_0^N
     = \int_0^N \langle \mathcal{L} \rangle_N' dN'.
\end{equation}
Reminding that $F(N=0) = 0$, being the free energy an extensive quantity, we have
\begin{equation}
   \frac{F}{N} + \frac{1}{\beta} \log2+ \frac{\beta}{4} 
     = \frac{1}{ \beta N } \int_0^N \langle \mathcal{L} \rangle_N' dN'.
\end{equation}
Subtracting $ \mathcal{L}\left[\overline{q}(x)\right]$, we get
\begin{equation}
\label{eq:ourobservable}
  f - f_\infty  = \frac{1}{ \beta N }\int_0^N \langle \Delta \mathcal{L} \rangle_N dN,
\end{equation}
where $\Delta \mathcal{L}\left[q(x)\right] = \mathcal{L}\left[q(x)\right] - \mathcal{L}\left[\overline{q}(x)\right]$, and $f_\infty $ is defined in (\ref{eq:f_infty}). 
We note here that the quantity we are interested in is the variation of the effective action between the finite size system and the infinite one. 
Since we believe that the left hand side of (\ref{eq:ourobservable}) behaves like a power of $N$, we expect the same also for $ \int_0^N \langle \Delta \mathcal{L} \rangle_N dN $. 
We have:
\begin{equation}
     \frac{1}{N} \int_0^N \langle \Delta \mathcal{L} \rangle_{N'} dN \sim N^\gamma \iff    \langle \Delta \mathcal{L} \rangle_{N} \sim N^{\gamma}.
\end{equation}
Hence,
\begin{equation}
\label{eq:free_energy_expert}
   f \sim \langle \Delta \mathcal{L} \rangle_N \sim N^{\gamma}.
\end{equation}
We easily notice that the exponent $\alpha$ we are looking for is exactly the opposite of $\gamma$.
Therefore,
\begin{equation}
\langle \Delta \mathcal{L} \rangle_N \sim N^{-\alpha},
\end{equation}
as announced in (\ref{eq:free_energy_double_expert}).
So, we can estimate $\alpha$ by studying how the observable $\Delta \mathcal{L}$ scales with $N$.
Since we are only interested in the behavior of $f$ as a function of $N$, (\ref{eq:free_energy_double_expert}) allows us to avoid ever calculating the normalization factor $Z$ in (\ref{eq:startnumerical}).

\printbibliography
\end{document}